\documentclass[twocolumn]{aastex63}

\usepackage{comment}
\usepackage{cases}
\usepackage{xcolor}

\received{April 15, 2020}
\revised{June 11, 2020}
\accepted{June 16, 2020}
\submitjournal{APJL}

\shorttitle{Morita Array [\ion{C}{1}]~$^3P_1$--$^3P_0$ and $^{12}$CO~$J$=4--3 Observations of NGC~6052}
\shortauthors{Michiyama et al.}
\graphicspath{{./}{figures/}}
\begin{document}

\title{Discovery of a [C~I]-faint, CO-bright Galaxy: ALMA Observations of the Merging Galaxy NGC~6052}

\correspondingauthor{Tomonari Michiyama}
\email{t.michiyama.astr@gmail.com}

\author[0000-0003-2475-7983]{Tomonari Michiyama}
\affiliation{Kavli Institute for Astronomy and Astrophysics, Peking University, 5 Yiheyuan Road, Haidian District, Beijing 100871, P.R.China}
%\affiliation{Department of Astronomy, School of Physics, Peking University, Beijing 100871, China}

\author{Junko Ueda}
\affiliation{National Astronomical Observatory of Japan, National Institutes of Natural Sciences, 2-21-1 Osawa, Mitaka, Tokyo, 181-8588}

\author{Ken-ichi Tadaki}
\affiliation{National Astronomical Observatory of Japan, National Institutes of Natural Sciences, 2-21-1 Osawa, Mitaka, Tokyo, 181-8588}

\author{Alberto Bolatto}
\affiliation{Department of Astronomy and Laboratory for Millimeter-Wave Astronomy, University of Maryland, College Park, MD 20742, USA}

\author{Juan Molina}
\affiliation{Kavli Institute for Astronomy and Astrophysics, Peking University, 5 Yiheyuan Road, Haidian District, Beijing 100871, P.R.China}
%\affiliation{Department of Astronomy, School of Physics, Peking University, Beijing 100871, China}

\author{Toshiki Saito}
\affiliation{Max-Planck-Institut f\"ur Astronomie, K\"onigstuhl 17, 69117 Heidelberg, Germany}

\author{Takuji Yamashita}
\affiliation{National Astronomical Observatory of Japan, National Institutes of Natural Sciences, 2-21-1 Osawa, Mitaka, Tokyo, 181-8588}
\affiliation{Research Center for Space and Cosmic Evolution, Ehime University, 2-5 Bunkyo-cho, Matsuyama, Ehime 790-8577, Japan}

\author{Ming-Yang Zhuang}
\affiliation{Kavli Institute for Astronomy and Astrophysics, Peking University, 5 Yiheyuan Road, Haidian District, Beijing 100871, P.R.China}
\affiliation{Department of Astronomy, School of Physics, Peking University, Beijing 100871, China}

\author{Kouichiro Nakanishi}
\affiliation{National Astronomical Observatory of Japan, National Institutes of Natural Sciences, 2-21-1 Osawa, Mitaka, Tokyo, 181-8588}
\affiliation{Department of Astronomical Science, The Graduate University for Advanced Studies, SOKENDAI, 2-21-1 Osawa, Mitaka, Tokyo 181-8588}

\author{Daisuke Iono}
\affiliation{National Astronomical Observatory of Japan, National Institutes of Natural Sciences, 2-21-1 Osawa, Mitaka, Tokyo, 181-8588}
\affiliation{Department of Astronomical Science, The Graduate University for Advanced Studies, SOKENDAI, 2-21-1 Osawa, Mitaka, Tokyo 181-8588}

\author{Ran Wang}
\affiliation{Kavli Institute for Astronomy and Astrophysics, Peking University, 5 Yiheyuan Road, Haidian District, Beijing 100871, P.R.China}
\affiliation{Department of Astronomy, School of Physics, Peking University, Beijing 100871, China}

\author{Luis C. Ho}
\affiliation{Kavli Institute for Astronomy and Astrophysics, Peking University, 5 Yiheyuan Road, Haidian District, Beijing 100871, P.R.China}
\affiliation{Department of Astronomy, School of Physics, Peking University, Beijing 100871, China}



\begin{abstract}
We report sensitive [\ion{C}{1}]~$^3P_1$--$^3P_0$ and $^{12}$CO~$J$=4--3 observations of the nearby merging galaxy NGC~6052 using the Morita (Atacama Compact) Array of ALMA.
We detect $^{12}$CO~$J$=4--3 toward the northern part of NGC~6052, but [\ion{C}{1}]~$^3P_1$--$^3P_0$ is not detected with a [\ion{C}{1}]~$^3P_1$--$^3P_0$ to  $^{12}$CO~$J$=4--3 line luminosity ratio of$~\lesssim0.07$.
%NGC~6052 is the most [C~I]-poor nearby galaxy known.
According to models of photodissociation regions, the unusual weakness of [\ion{C}{1}]~$^3P_1$--$^3P_0$ relative to  $^{12}$CO~$J$=4--3 can be explained if the interstellar medium has a hydrogen density larger than $10^5\,{\rm cm}^{-3}$, conditions that might arise naturally in the ongoing merging process in NGC~6052. 
Its [C~I]~$^3P_1$--$^3P_0$ emission is also weaker than expected given the molecular gas mass inferred from previous measurements of  $^{12}$CO~$J$=1--0  and  $^{12}$CO~$J$=2--1.
This suggests that [C~I]~$^3P_1$--$^3P_0$ may not be a reliable tracer of molecular gas mass in this galaxy.
NGC~6052 is a unique laboratory to investigate how the merger process impacts the molecular gas distribution.
%Our study suggests that [C~I]-poor, CO-rich systems may have been missed in previous observational and theoretical investigations, 
%caution that [C~I] may not always be a reliable tracer of molecular gas mass in galaxies.
\end{abstract}



\section{Introduction} \label{sec:intro}
The atomic carbon emission line with the lower forbidden $^3P$ fine structure 
([\ion{C}{1}]~$^3P_1$--$^3P_0$, hereafter [\ion{C}{1}]~(1--0)) 
has a frequency of 492~GHz ($\lambda\simeq609$~$\mu$m). 
In the classical framework of photodissociation region (PDR) theory \citep[e.g.,][]{Tielens_1985}, 
the ultraviolet (UV) photons from young massive stars control 
the properties of the interstellar medium (ISM) and [\ion{C}{1}] emission mostly arises from a narrow transition layer between [\ion{C}{2}] and CO. 
On the other hand, [\ion{C}{1}] observations 
for Galactic molecular clouds (0.1-1~pc scales) 
frequently show spatial coincidence between [\ion{C}{1}] and CO emission lines and brightness temperatures very similar to that of $^{13}$CO~$J$=1--0
\citep[e.g.,][]{Ojha_2001, Oka_2001, Ikeda_2002, Kramer_2008, Shimajiri_2013, Burton_2015}. 
This suggests that [\ion{C}{1}] emission can be used to trace the bulk molecular gas mass in clouds.
\citet{Papa_2004_theory} proposed the ubiquitous distribution of atomic carbon 
throughout a typical Giant Molecular Cloud 
is due to the combination of dynamics and the non-equilibrium chemistry processes, and \citet{Papa_2004_observation} showed agreement between molecular gas masses estimated from [\ion{C}{1}] and $^{12}$CO emission lines in Arp~220 and NGC~6240. 

An increasing number of [\ion{C}{1}] emission line observations of extra-galactic sources have been reported. 
The Spectral and Photometric Imaging Receiver Fourier Transform Spectrometer (SPIRE/FTS) 
mounted on the {\rm Hershel Space Observatory} measured 
[\ion{C}{1}] luminosities for Ultra/ Luminous Infrared Galaxies (U/LIRGs) \citep{Kame_2016, Lu_2017}. 
The operation of high-frequency receivers (e.g., Band 8 and 10 receivers) 
on the Atacama Large Millimeter/submillimeter Array (ALMA) 
has also enabled investigations of the $\sim10$~pc resolution distribution of atomic carbon 
in the nucleus ($<1$kpc) of nearby galaxies 
\citep[e.g.,][]{Krips_2016, Izumi_2018, Miyamoto_2018, Salak_2019} 
and $\sim100$~pc scale maps of U/LIRGs 
such as NGC~6240 \citep{Cicone_2018} and
IRAS F18293-3413 (Saito et al., in prep).
In addition, there is increasing availability of 
[\ion{C}{1}] measurements of bright high-z galaxies, 
such as gravitational lensing galaxies and submillimter galaxies (SMGs),
that have been conducted using millimeter/submillimeter interferometers 
\citep[e.g.,][]{Walter_2011, Bothwell_2017, Tadaki_2018, Valentino_2018, Valentino_2020}.

The relation between [\ion{C}{1}] and CO has been investigated using these data. 
\citet{Jiao_2017,Jiao_2019} show a typical integrated brightness temperature ratio (i.e., units in [K~km~s$^{-1}$~pc$^2$]) of [\ion{C}{1}]~(1--0)/CO~(1--0)$\sim0.2$ 
and possible enhancement of [\ion{C}{1}] abundance in U/LIRGs with respect to normal star-forming galaxies.
\citet{Salak_2019} find variations of the [\ion{C}{1}]~(1--0)/CO~(1--0) ratio in 30–50~pc resolution data in the starburst galaxy NGC~1808, suggesting it is due to different excitation conditions and/or carbon abundance in different regions.
\citet{Valentino_2018, Valentino_2020} shows a line luminosity ratio of [\ion{C}{1}]~(1--0)/CO~(4--3)$\sim0.1-3$  
(in units of [$L_{\odot}$]) in $z\sim1-2$ main sequence galaxies, 
suggesting this is driven by changes of the density in the range 
$n\sim10^3$--$10^5$~[cm$^{-3}$] 
based on a simple PDR model \citep[see also ][]{Bothwell_2017}.

The determination of the conversion factor from [\ion{C}{1}] luminosity to molecular gas mass is a current research topic 
\citep{Israel_2015, Crocker_2019, Heintz_2020}. 
Recent theoretical models posit that
atomic carbon is well-mixed with H$_2$ in several environments, 
such as metal-poor and/or sources with high-cosmic ray fluxes \citep[e.g.,][]{Papa_2018}. 
In such environments [\ion{C}{1}]-rich/CO-poor gas may be very abundant, 
suggesting that [\ion{C}{1}] could be better than CO as a molecular gas mass tracer.

Summarizing the previous work, [\ion{C}{1}] emission is expected to be an indicator of the cold molecular gas but its strength may depend on various parameters. To investigate it further it is necessary to collect a wider range of observations.
The Morita Array (the official name of the Atacama Compact Array of 7~m diameter dishes) of ALMA enables us to investigate 
nearby galaxies with [\ion{C}{1}] fluxes that were 
too low to be detectable with SPIRE/FTS. 
In this letter, we report the non-detection of [\ion{C}{1}]~(1--0) 
and the detection of CO~(4--3) in NGC~6052 using the Morita Array. 
%Such [\ion{C}{1}]-poor/CO-rich galaxies  are not well documented in previous surveys, and their existence suggests that [\ion{C}{1}] might not always be a reliable tracer of molecular gas mass in galaxies.

\section{NGC~6052} \label{sec:N6052} 
\begin{figure*}[!htbp]
\begin{center}
\includegraphics[width=18cm]{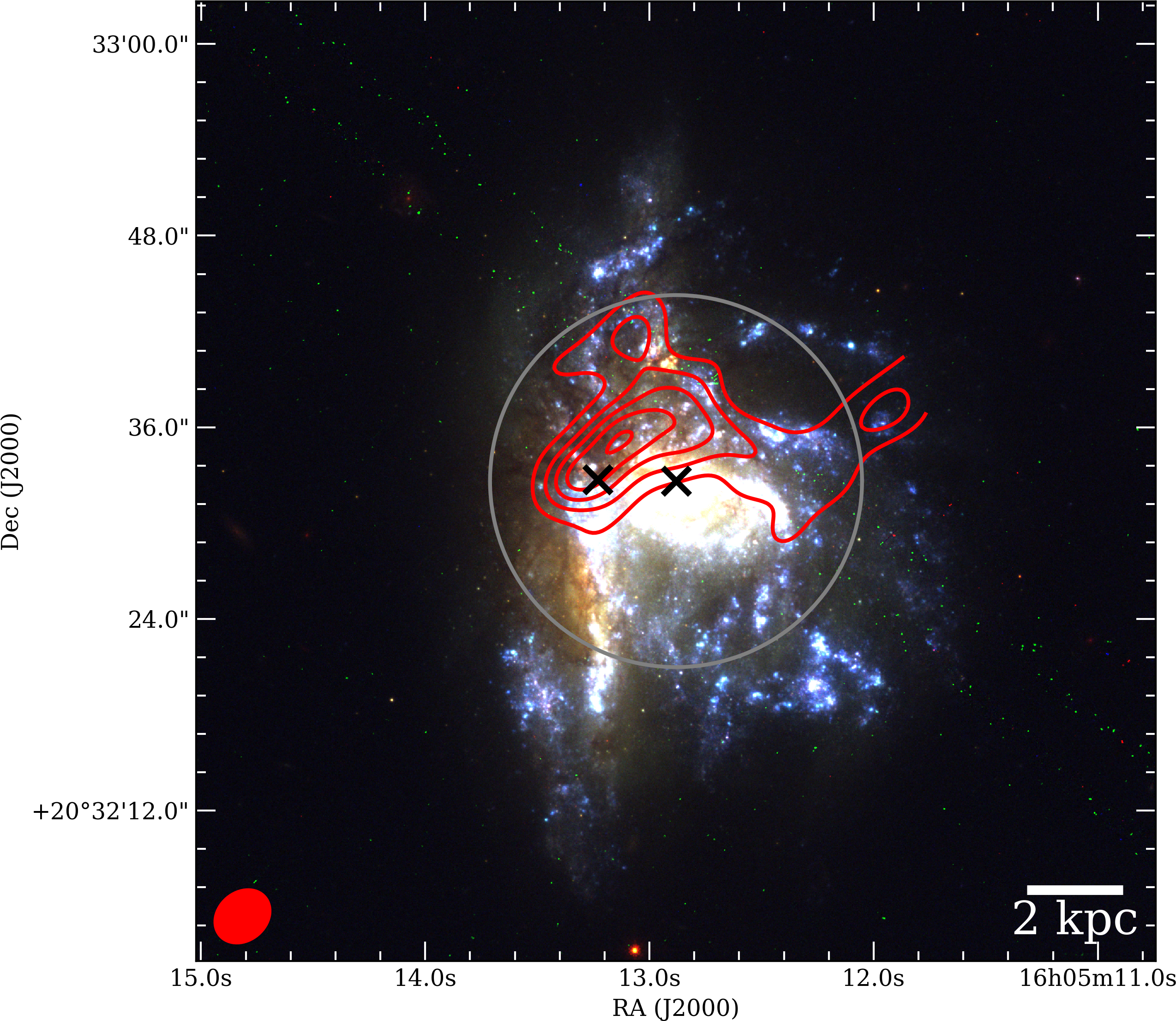}
\caption{
The three-color composite image of NGC~6052
from HST observations using the WFC3-F814W/F555W/F336W filters. 
The red contours show the CO~(4-3) spatial distribution 
obtained from the Morita Array observation 
at 3, 6, 9, 12, and 15 sigma levels with an achieved r.m.s of 3.6\,Jy\,km\,s$^{-1}$\,beam$^{-1}$.
The grey circle shows the 
FoV (full width at half maximum of the primary beam
$\sim23\farcs5$ at the CO~(4--3) sky frequency).
The West and East nuclei are shown as X symbols. 
The pointing center is indicated by the western black X symbol 
($16^{\rm{h}} 05^{\rm{m}} 12\fs88$, $+20\degr 32\arcmin 32\farcs61$ in ICRS frame).
The position of the Western and Eastern nucleus is based on NED.
The synthesized beam ($3\farcs8 \times 3\farcs1$) and the spatial scale are shown at the left- and right-bottom corners, respectively.
\label{fig:HST}}
\end{center}
\end{figure*}

NGC~6052 (Arp 209, UGC 10182, VV 86, Mrk 297) is a system of two colliding galaxies  (east and west components)
at a redshift of $z=0.01581$\footnote{Based on NASA/IPAC Extragalactic Database (NED).} (the corresponding luminosity distance is $D_L=70.8$~Mpc, and the scale is $1\farcs0\sim333$~pc)
\footnote{Using the cosmology by Planck Collaboration Paper XIII, $H_0=67.7$ and $\Omega_m=0.307$ \citep{Planck_2016}.}.
The systemic velocity of $V_{\rm sys} = 4666$ km~s$^{-1}$ is calculated based on the radio velocity convention.
The background of Figure~\ref{fig:HST} 
is a three-color composite image of NGC~6052 
obtained by the Wide Field Camera 3 (WFC3) onboard the NASA/ESA 
{\rm Hubble~Space~Telescope}\footnote{The {\rm WFC3} data were obtained from Hubble legacy archive 
(\url{https://hla.stsci.edu/hlaview.html}) with the dataset name of
``hst$\textunderscore$14066$\textunderscore$05$\textunderscore$wfc3$\textunderscore$uvis$\textunderscore$f814w$\textunderscore$f555w$\textunderscore$f336w". }.
The positions of the West and East nuclei are shown by X symbols.
Modeling by  N-body simulations suggests that the evolutionary phase of this colliding system is about  $\sim150$\,Myr after the first impact \citep{Taniguchi_1991}.
The total infrared luminosity (8-1000~$\mu$m, $L_{\rm TIR}$) of NGC~6052 is
$\sim10^{11.0}~L_{\odot}$ based on {\rm Infrared Astronomical Satellite (IRAS)} photometry \citep{Sanders_2003}, 
which corresponds to a star formation rate (SFR) of SFR$\sim18$~M$_\odot$~yr$^{-1}$ using the calibration by \citet{Mo_2010}.
The metallicity of NGC 6052 has been calculated in literature, but the values vary depending on the method: e.g., $12+\log({\rm [O]/[H])}$ = 
8.85 \citep{Sage_1993}
8.65 \citep{James_2002}, 
8.34 \citep{Shi_2005}, and
8.22-8.80 \citep{Rupke_2008}.
For the sake of simplicity, we assume the Milky Way metallicity for PDR modeling in SECTION \ref{PDR}.

\section{Morita Array Observation} \label{sec:ACA}
\begin{deluxetable*}{cccccc}
%\tablenum{1}
\tablecaption{Morita Array observation summary for NGC~6052 \label{tab:ACA}}
\tablewidth{0pt}
\tablehead{
\colhead{Line} & \colhead{frequency (rest/sky)} & \colhead{Synthesized beam size} & \colhead{r.m.s} & \colhead{Velocity Range} & \colhead{FWHM}\\
\colhead{} & \colhead{GHz} & \colhead{arcsec} & \colhead{mJy~beam$^{-1}$} & \colhead{km~s$^{-1}$} & \colhead{km~s$^{-1}$}
}
\decimalcolnumbers
\startdata
[\ion{C}{1}]~(1--0) & 492.16 / 484.50 & $3\farcs1\times2\farcs8$ & 61.5 & [-72, 63] & -- \\
CO~(4--3) & 461.04 / 453.86 & $3\farcs8\times3\farcs1$ & 46.8 & [-72, 63]  & 96
\enddata
\tablecomments{(1) Line name. 
(2) The rest-frame and sky frequencies of the emission lines. 
(3) Synthesized beam size. 
(4) The r.m.s level of the data cube 
with the velocity resolution of 15~km~s$^{-1}$. 
(5) The velocity range for the velocity-integrated intensity map. 
The values are the relative velocity with respect to the systematic velocity of 4666\,km\,s$^{-1}$.
(6) The FWHM of the emission line.}
\end{deluxetable*}
As a part of our ALMA Cycle 6 project (2018.1.00994.S), 
we carried out observations of NGC~6052  
for [\ion{C}{1}]~(1--0) and CO~(4--3) 
using a single-pointing 
on 2019 April 21 and 2019 June 3, respectively. 
The pointing center is the West nucleus 
and the half-power field of view (FoV) diameter is 23$\farcs$5 at the frequency of CO~(4--3) (Figure~\ref{fig:HST}), with a maximum recoverable scale (MRS) of 13$\farcs$9 
($\sim4.6$~kpc for NGC~6052).
The corresponding figures for [\ion{C}{1}]~(1--0) are 7\% smaller.
The calibrated visibility data were obtained 
by running the provided calibration scripts 
using $\tt{CASA}$ (version: 5.4.0-70) \citep{McMullin_2007}. 
The imaging was performed using $\tt{tclean}$ in $\tt{CASA}$
with a velocity resolution of 15~km~s$^{-1}$, cell size of 0$\farcs$2, image size of 200 pixel, 
and Briggs weighting with a robust parameter of 0.5.
The clean masks for each channel were determined manually.
The synthesized beam size and 
sensitivity at the velocity resolution of 15~km~s$^{-1}$ 
are shown in columns (3) and (4) of Table~\ref{tab:ACA}, respectively.

\begin{figure*}[!htbp]
\begin{center}
\includegraphics[width=18cm]{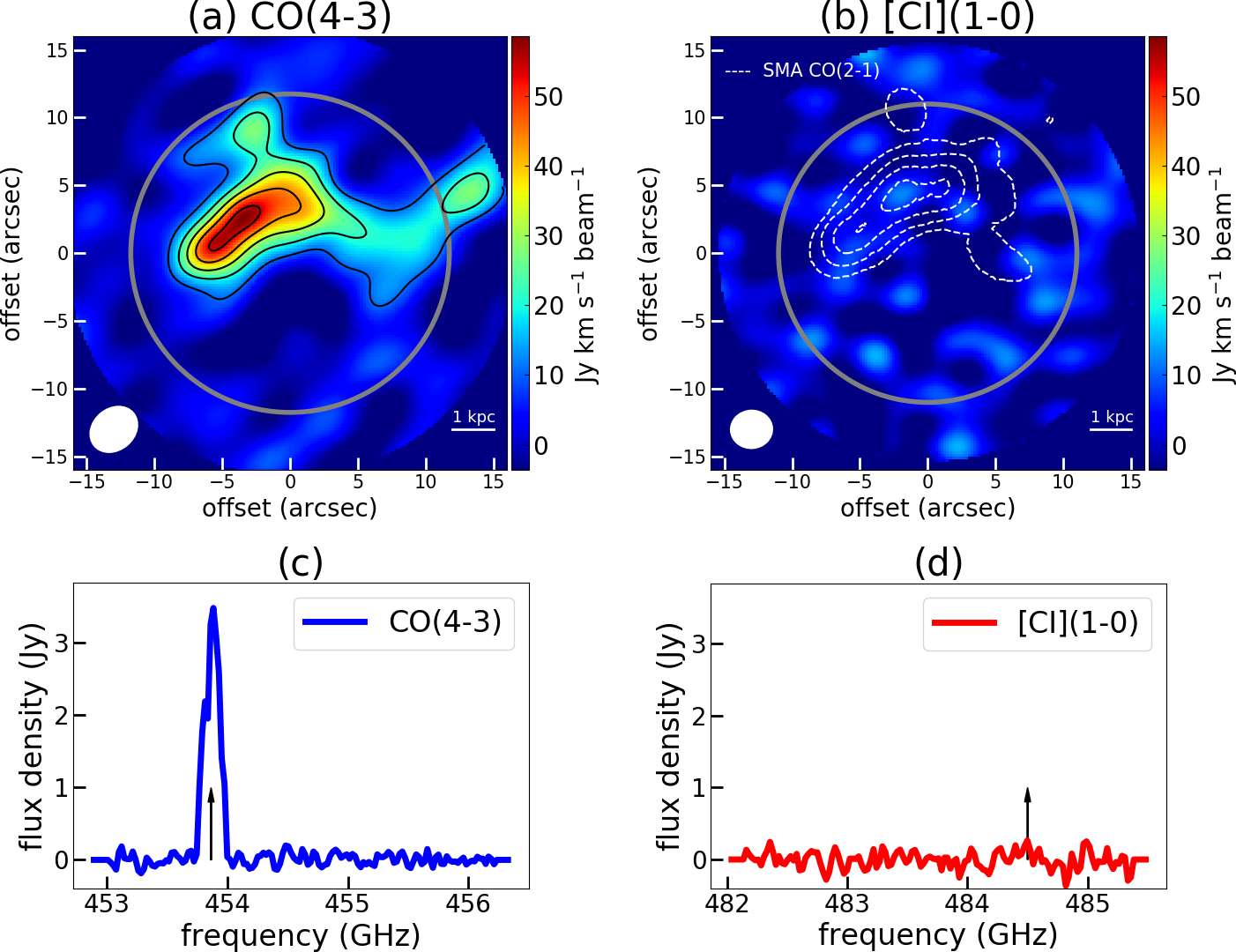}
\caption{(top) Velocity-integrated intensity maps 
of the (a) CO~(4--3) and (b) [\ion{C}{1}]~(1--0) emission lines. 
These were created by integrating
the data cubes within the [-72, 63] km\,s$^{-1}$ velocity range 
(the systemic velocity is 4666~km~s$^{-1}$) 
before applying the primary beam correction.
In the CO~(4--3) integrated intensity map, 
the black contours represent the same signal-to-noise levels as in Figure~\ref{fig:HST}. 
In the [\ion{C}{1}]~(1--0) map, 
the synthesized beam size is $3\farcs1 \times 2\farcs8$ 
and the achieved r.m.s level is 5.6\,Jy\,km\,s$^{-1}$\,beam$^{-1}$.
The synthesized beam and the spatial scale are shown at the left- and right-bottom corners, respectively.
The white dashed lines show the CO~(2-1) distribution obtained by SMA with the contour levels of (1.7, 8.5, 15.3, 22.1)\,Jy\,km\,s$^{-1}$\,beam$^{-1}$ \citep{Ueda_2014}.
(bottom)
The spectra integrated over the primary beam area. 
The velocity resolution is 15\,km\,$\rm s^{-1}$ in both spectra. 
The r.m.s of the spectra is 0.07~Jy and 0.1~Jy for (c) and (d), respectively.
The black arrows indicate 
the frequencies of
the CO~(4--3) and [\ion{C}{1}]~(1--0) corresponding to the systemic velocity.
 \label{fig:mom0}}
\end{center}
\end{figure*}

The intensity map integrated within the velocity range of [-72, 63]~km~s$^{-1}$ 
is presented in Figure~\ref{fig:mom0}.
The CO~(4--3) emission line is detected in the northern part of the galaxy, 
and it is offset from the optical peak.
The CO~(4--3) distribution is similar to CO~(2--1) obtained by Submillimeter Array (SMA) \citep{Ueda_2014} (Figure~\ref{fig:mom0}(b) white dashed contour).
[\ion{C}{1}]~(1--0) emission is not detected 
within the FoV, particularly it is undetected in the region detected on the CO~(4--3) and CO~(2--1) peaks.
We measure a velocity-integrated CO~(4--3) flux of 
$S_{\rm CO(4-3)}\Delta v = 414\pm124$~Jy~km~s$^{-1}$
by integrating the spectrum over the entire FoV after correcting for the primary beam response (Figure~\ref{fig:mom0}(c), the velocity range is shown in Table~\ref{tab:ACA}). 
In this letter, we adopt 30~\% as a conservative systematic uncertainly error, but this is not critical for our conclusions.
We note that the error can be estimated from the r.m.s level of the channel map (7~Jy~km~s$^{-1}$) with a systematic uncertainty associated with the expected absolute flux accuracy of 10~\% for ALMA Band 8\footnote{\url{https://almascience.nao.ac.jp/documents-and-tools/cycle6/alma-proposers-guide}}. 
However, the actual performance of the flux calibration has not been explored for the ACA stand-alone mode in Band~8 by the ALMA observatory.

We obtain a $3\sigma$ upper limit for the [\ion{C}{1}]~(1--0) velocity integrated intensity using
\begin{eqnarray}
 S_{\rm line}{\rm d} v&<&3\sigma_{\rm ch} D_v \sqrt{\frac{N_{\rm ch}N_{\rm box}}{ N_{\rm beam}}} \nonumber \\
 &=& 28~{\rm Jy~km~s^{-1}},
\end{eqnarray}
where 
$\sigma_{\rm ch}=61.5$~[mJy~beam$^{-1}$] is the r.m.s revel of the channel map (Table~\ref{tab:ACA}), 
$D_v=15$~km~s$^{-1}$ is the velocity resolution,
$N_{\rm ch} =FWHM/D_v$, 
$FWHM=96$~[km~s$^{-1}$] assuming the same value with CO~(4--3) line shape (Figure~\ref{fig:mom0}(c)),
$N_{\rm box} = 3990$ (the number of pixel where CO~(4--3) is detected inside the three sigma contour of Figure~\ref{fig:mom0}(b)),
$N_{\rm beam} = 247$ is the number of pixel for the synthesized beam, following the equation from \citet{Hainline_2004}. 

The velocity integrated line fluxes 
and line luminosities including CO~(1--0) and CO(2--1) values 
from literature \citep{Albrecht_2007, Ueda_2014}
are summarized in Table~\ref{tab:line}. 
The luminosities for the lines are measured using 
\begin{eqnarray}\label{L1}
L_{\rm line}&&[L_{\odot}]  \nonumber \\
&& =1.04\times10^{-3}~S_{\rm line}\Delta v~ \nu_{\rm rest} ~(1+z)^{-1}~D_{\rm L}^2
\end{eqnarray}
and
\begin{eqnarray}\label{L2}
L'_{\rm line}&&[{\rm K~km~s^{-1}~pc^2}]  \nonumber \\
&&= 3.25\times10^{7}~S_{\rm line}\Delta v~ \nu_{\rm obs}^{-2} ~(1+z)^{-3}~D_{\rm L}^2,
\end{eqnarray}
where $S_{\rm line}\Delta v$ is the velocity-integrated flux in Jy~km~s$^{-1}$, 
$\nu_{\rm rest}$ is the rest frequency of the observing line emission in GHz, 
$z$ is redshift, $D_{\rm L}$ is the luminosity distance in Mpc, 
and $ \nu_{\rm obs}=\nu_{\rm rest}/(1+z)$ is \added{the} sky frequency 
for the observing line emission in GHz \citep{Solomon_2005}. 
We use $L'_{\rm line}$ to investigate the molecular gas mass (SECTION~\ref{gas_mass}) 
and $L_{\rm line}$ to investigate the luminosity ratios for a PDR modeling (SECTION~\ref{PDR}).

\begin{deluxetable*}{cccccc}
%\tablenum{1}
\tablecaption{Line properties of NGC~6052 \label{tab:line}}
\tablewidth{0pt}
\tablehead{
\colhead{Line} & \colhead{$S_{\rm line}$d$v$} & \colhead{$L'_{\rm line}$} & \colhead{$L_{\rm line}$} & \colhead{telescope} & \colhead{reference} \\
\colhead{} & \colhead{Jy~km~s$^{-1}$} & \colhead{K km~s$^{-1}$~pc$^2$} & \colhead{$L_{\odot}$}
}
\decimalcolnumbers
\startdata
$[$\ion{C}{1}$]$~(1--0) & $<28$ & $<2\times10^7$  & $<7\times10^4$ & Morita Array & this work \\
CO~(1--0)  & 82 & $9.8\times10^8$  & $4.7\times10^4$ & IRAM~30m &\citet{Albrecht_2007} \\
CO~(2--1)  & $162\pm32$ & $(4.9\pm1.0)\times10^8$  & $(1.9\pm0.4)\times10^5$ & SMA &\citet{Ueda_2014} \\
CO~(4--3)& $414\pm124$ & $(3.1\pm0.9)\times10^8$  & $(9.8\pm2.9)\times10^5$ & Morita Array &  this work \\
%$[$\ion{C}{2}$]_{158}$ & $3.8\times10^4$ & 
%$1.7\times10^9$  & $3.7\times10^8$ &PACS/\rm{Hershel} &\citet{Diaz-Santos_2017} \\
\enddata
\tablecomments{(1) Line name. 
(2) The velocity integrated flux. We use the Jy to K conversion of 4.95 to derive the $S_{\rm line}$d$v$ of CO~(1--0) from the integrated intensity reported as 16.4 [K~km~s$^{-1}$] in the 24\arcsec\ single-dish beam.
(3) and (4) The line luminosity calculated by equations \ref{L1} and \ref{L2}. 
(5) The telescope name.
(6) The reference literature for the line properties.
}
\end{deluxetable*}

\section{Discussion}
\subsection{Photodissociation Regions}\label{PDR}
\begin{figure*}[!htbp]
\begin{center}
\includegraphics[width=18.5cm]{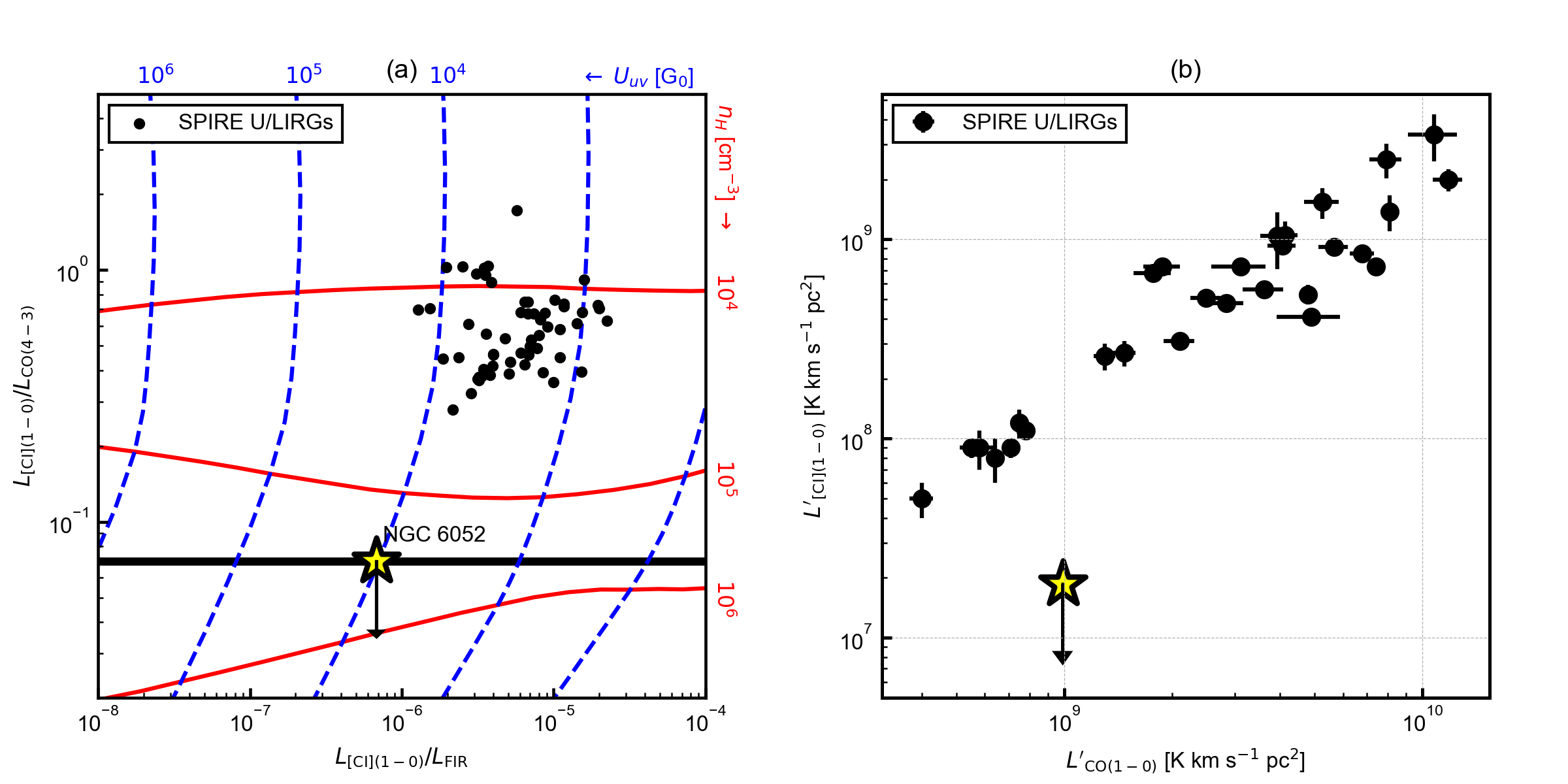}
\caption{
(a) The $L_{\rm [CI](1-0)}/L_{\rm CO(4-3)}$ versus $L_{\rm [CI](1-0)}/L_{\rm FIR}$ for PDRT calculations and observations. 
The red solid and blue dashed lines indicate the theoretical tracks for constant
$n_{\rm H}$~[cm$^{-3}$] and $U_{\rm uv}$~[$G_0$], respectively.
The black points indicate the nearby U/LIRG sample detected by SPIRE/FTS \citep{Kame_2016}.
The yellow star shows the [\ion{C}{1}]~(1--0) upper limit of NGC~6052 presented in this project.
The $L_{\rm [CI](1-0)}/L_{\rm TIR)}$ is calculated without any aperture correction, indicating that this figure does not constrain the $U_{\rm u,v}$.
(b) The [\ion{C}{1}]~(1--0) line luminosity versus the CO~(1--0) line luminosity for U/LIRG galaxies. The symbols are the same comparison sample as (a). We note that CO~(1--0) luminosity is obtained by single dish but the upper limit of [\ion{C}{1}]~(1--0) luminosity is obtained by array observation (the missing flux is not corrected).
\label{fig:Discussion}}
\end{center}
\end{figure*}

In this section, we explain low $L_{\rm [CI](1-0)}/L_{\rm CO(4-3)}$ of $< 0.07$, which is not well documented in previous observations such as \citet{Israel_2002, Valentino_2020}.
%Luminosity ratios such as $L_{\rm [CI](1-0)}/L_{\rm CO(4-3)}$ and $L_{\rm [CI](1-0)}/L_{\rm TIR}$ can be used to constrain the physical parameters of PDRs. \citep[e.g., ][]{Alaghband-Zadeh_2013,Bothwell_2017}.
We use the Photodissociation Region Toolbox 
({\tt PDRT}\footnote{\url{http://dustem.astro.umd.edu/pdrt/}}) 
\citep{Kaufman_1999, Kaufman_2006, Pound_2008} 
to investigate density ($n_{\rm H}$) and
incident FUV radiation field ($U_{uv}$ corresponding to photons with 6~eV$\leq h\nu <$ 13.6~eV) 
in units of the average interstellar radiation field in the vicinity of the Sun
($G_0 =1.6\times10^{-3}$ergs~cm$^{-2}$~s$^{-1}$)
\citep{Habing_1968}.
The modeling predicts line intensities 
for combinations of $n_{\rm H}$ and $U_{uv}$ 
by self-consistently solving for chemical processes, radiation transfer, and thermal balance.
Figure \ref{fig:Discussion}(a) shows the predictions for 
$L_{\rm [CI](1-0)}/L_{\rm CO(4-3)}$ versus $L_{\rm [CI](1-0)}/L_{\rm TIR}$ together with the data 
for nearby U/LIRGs observed by SPIRE/FTS \citep{Kame_2016}, 
and the upper limits we obtain for NGC~6052.
The red solid and blue dashed lines indicate 
the tracks for constant $n_{\rm H}$ and $U_{\rm uv}$, respectively.
According to the plane-parallel PDR model,
the UV radiation cannot penetrate deeply into the molecular medium 
if the region is dense, 
yielding the narrow [\ion{C}{1}] layer and hence low  [\ion{C}{1}] flux. 
The upper limit for NGC~6052 ($L_{\rm [CI](1-0)}/L_{\rm CO(4-3)}<0.07$) 
suggests a dense PDR ($n_{\rm H}>10^5$~[cm$^{-3}$]), 
which is denser than those seen in the comparison galaxies \citep[e.g.,][]{Bothwell_2017, Valentino_2018, Valentino_2020}. 

A possible explanation of high densities is compression of the ISM 
by the ongoing merger 
because the region associated with the CO~(4--3) emission
corresponds to the collision front 
based on the merger model calculation \citep{Taniguchi_1991}.
Furthermore, plane-parallel model might be suitable for young off-nuclear starburst regions that have not had time to erode the medium.
The age of the stellar population at CO detected region is estimated to be $\sim5.5$~Myr by modeling of the mid-infrared atomic lines \citep{Whelan_2007}, suggesting the presence of young off-nuclear starbursts 
triggered by the ongoing merger.  
The dense PDR regions associated with young off-nuclear starburst activities might be a hint to understand the [C I]-poor, CO-rich region in NGC~6052.

The PDRT calculations suggest that the small $L_{\rm [CI](1-0)}/L_{\rm TIR}$ is indicative of a high $U_{\rm uv}$.
However, it is not possible to obtain $U_{\rm uv}$ from the $L_{\rm [CI](1-0)}/L_{\rm TIR}$ obtained from our observations.
This is because the ratio of the TIR luminosity in the observable region of the Morita Array to the galaxy integrated total TIR luminosity (determined by IRAS measurement) is unknown.
Instead, we can access the galaxy integrated value of ([\ion{O}{1}]$_{63}$ + [\ion{C}{2}]$_{158}$)/TIR\footnote{The flux of [\ion{O}{1}]$_{63}$ and [\ion{C}{2}]$_{158}$ is 82 and 138 [W~m$^{-1}$] \citep{Diaz-Santos_2017} and the TIR flux density of $6.5\times10^{-13}$ [W~m$-1$] is calculated from IRAS flux density measured by \citep{Sanders_2003}.} that can also characterize the PDR. The ratio suggests $U_{\rm uv}=4-5$. \citep{Diaz-Santos_2017}.

\subsection{Molecular gas mass}\label{gas_mass}
In this section, we investigate whether [\ion{C}{1}]~(1--0) can trace molecular gas mass inferred from previous CO~(1--0) measurements.  
\citet{Albrecht_2007} reported the CO~(1--0) integrated intensity of 82 K km~s$^{-1}$  within the 24\arcsec\ single-dish beam. The expected molecular mass is 
$M_{\rm H2}^{\rm CO} = 8 \times 10^{8}~M_{\odot}$ assuming a CO~(1--0) to H$_2$ conversion factor $\alpha_{\rm CO}$=0.8~$M_{\odot}$~(K~km~s$^{-1}$~pc$^2$)$^{-1}$ (where
$M_{\rm H2}^{\rm line} = \alpha_{\rm line} L'_{\rm line}$), one of the smallest values usually considered and often applied to ULIRGs \citep{Bolatto_2013}.
Using the upper limit of the [\ion{C}{1}]~(1--0) luminosity, we can estimate an upper limit to the implied molecular mass through the use of $\alpha_{\rm [CI]} = 7.3$~$M_{\odot}$~(K~km~s$^{-1}$~pc$^2$)$^{-1}$
\citep{Crocker_2019}.
The upper limit of molecular gas mass based on [\ion{C}{1}] is then
$M_{\rm H2}^{\rm [CI]} < 10^{8}~[M_{\odot}$].
This suggests that [\ion{C}{1}]~(1--0)/CO~(1--0) ratio is low as well as [\ion{C}{1}]~(1--0)/CO~(4--3) ratio.
Figure~\ref{fig:Discussion}(b) shows the the relation between CO~(1--0) and [\ion{C}{1}]~(1--0) luminosities for other nearby galaxies taken from \citet{Kame_2016}. The upper limit we measure for the [\ion{C}{1}]~(1--0)/CO~(1--0) ratio in NGC~6052 is located an order of magnitude below 
the global relation obtained in previous surveys \citep[e.g.,][]{Jiao_2017, Jiao_2019}.  
This suggests that [\ion{C}{1}]~(1--0) may not be a reliable tracer of molecular gas mass in NGC~6052.
Although both $\alpha_{\rm [CI]}$ and $\alpha_{\rm CO}$ depend on metallicity, such a [\ion{C}{1}]-poor region cannot be explained under the assumption of any metallicity
\citep[e.g.,][]{Bolatto_2013, Glover_2016, Heintz_2020}.
For example, at lower metallicities CO is would be more easily dissociated due to the lack of dust shielding, and consequently more carbon would be observed as [\ion{C}{1}].

An alternative explanation for the big difference between NGC~6052 and other systems in Figure~\ref{fig:Discussion} is resolving out spatially extended diffuse gas emission.
The emission from [\ion{C}{1}](1-0) and CO(1-0) are usually co-extensive, and both are likely to originate predominantly from extended, low excitation gas. 
On the other hand, higher-$J$ CO emission is likely dominated by compact, high excitation gas and the contribution from more diffuse gas.
Thus, the [\ion{C}{1}](1-0)/CO~(4--3) and [\ion{C}{1}]~(1--0)/CO~(1--0) ratio may represent properties of very different phases of the gas, and it is possible that the Morita Array resolved out most extended emission from [\ion{C}{1}]~(1--0).
%Thus, the [\ion{C}{1}](1-0)/CO~(4--3) ratio obtained by array observations may not trace the properties of global extended low excitation gas, and low [\ion{C}{1}]~(1--0)/CO~(1--0) ratio might be due to extended [\ion{C}{1}]~(1--0) structure that Morita Array cannot collect. 
In such a case, the large difference seen in Figure~\ref{fig:Discussion}(b) can be explained if $>$88~\% of the flux of [\ion{C}{1}]~(1--0) was missing due to its extended structure, possibly larger than the Maximum Recoverable Scale of $\sim4.3$~kpc.
While such large missing flux is not generally seen in other galaxies, follow-up mosaic mapping observations including 12m, Morita Array, and total power would be important to check the possibility of an extremely extended [\ion{C}{1}]~(1--0) distribution in NGC~6052. 
In either case, whether [\ion{C}{1}]-poor or with an extremely extended [\ion{C}{1}] distribution, NGC~6052 is a unique laboratory to investigate how the merger process impacts the use of [\ion{C}{1}] as a mass tracer.

\section{Summary} \label{sec:Summary}
We report [\ion{C}{1}]~(1--0) and CO~(4--3) observations 
of the nearby merging galaxy NGC~6052 using the ALMA Morita Array. 
We detect CO~(4--3) with high significance (signal-to-noise ratio of $>40$),  
but [\ion{C}{1}]~(1--0) is undetected to a stringent upper limit of 0.07 times the strength of the CO~(4--3) emission. 
Models of PDRs can explain 
the weakness of [\ion{C}{1}]
as the result of  gas densities that are unusually high
($n_{\rm H} > 10^5\,{\rm cm}^{-3}$), which might arise naturally 
in the collision front of the ongoing merger. 
In addition, [\ion{C}{1}](1--0) is far weaker than 
expected for the amount of molecular gas 
inferred from the existing measurements of CO~(1--0) and CO~(2--1) in NGC~6052.
This may suggest  [\ion{C}{1}]-poor, CO-rich system and/or extremely extended diffuse molecular gas distribution that is not well documented in literature.
%Our observation suggests that [C~I]-poor, CO-rich systems have been overlooked in previous observations, indicating that under some conditions [\ion{C}{1}] is not an effective tracer of the molecular gas mass in galaxies.

\acknowledgments
This work was supported by the National Science Foundation of China (11721303, 11991052) and the National Key R\&D Program of China (2016YFA0400702).
We are grateful to the anonymous referee for useful comments which helped the authors to improve the paper.
The authors appreciate Prof. Masami Ouchi (The University of Tokyo) and Prof. Takuya Hashimoto (Tsukuba University) who provided a comfortable and fruitful research environment during the COVID-19 pandemic time.
This paper makes use of the following ALMA data: ADS/JAO.ALMA $\#$2018.1 00994.
ALMA is a partnership of ESO (representing its member states), NSF (USA) and NINS (Japan), together with NRC (Canada), MOST and ASIAA (Taiwan), and KASI (Republic of Korea), in cooperation with the Republic of Chile. The Joint ALMA Observatory is operated by ESO, AUI/NRAO and NAOJ.

\clearpage
\appendix
\setcounter{section}{0}
\renewcommand{\thesection}{\Alph{section}}
\setcounter{figure}{0}
\renewcommand{\thefigure}{\Alph{section}.\arabic{figure}}
\setcounter{table}{0}
\renewcommand{\thetable}{\Alph{section}.\arabic{table}}
\section{Notes about SPIRE/FTS measurements}
\label{sec:Appendix}
\begin{figure}[!htbp]
\begin{center}
\includegraphics[width=18cm]{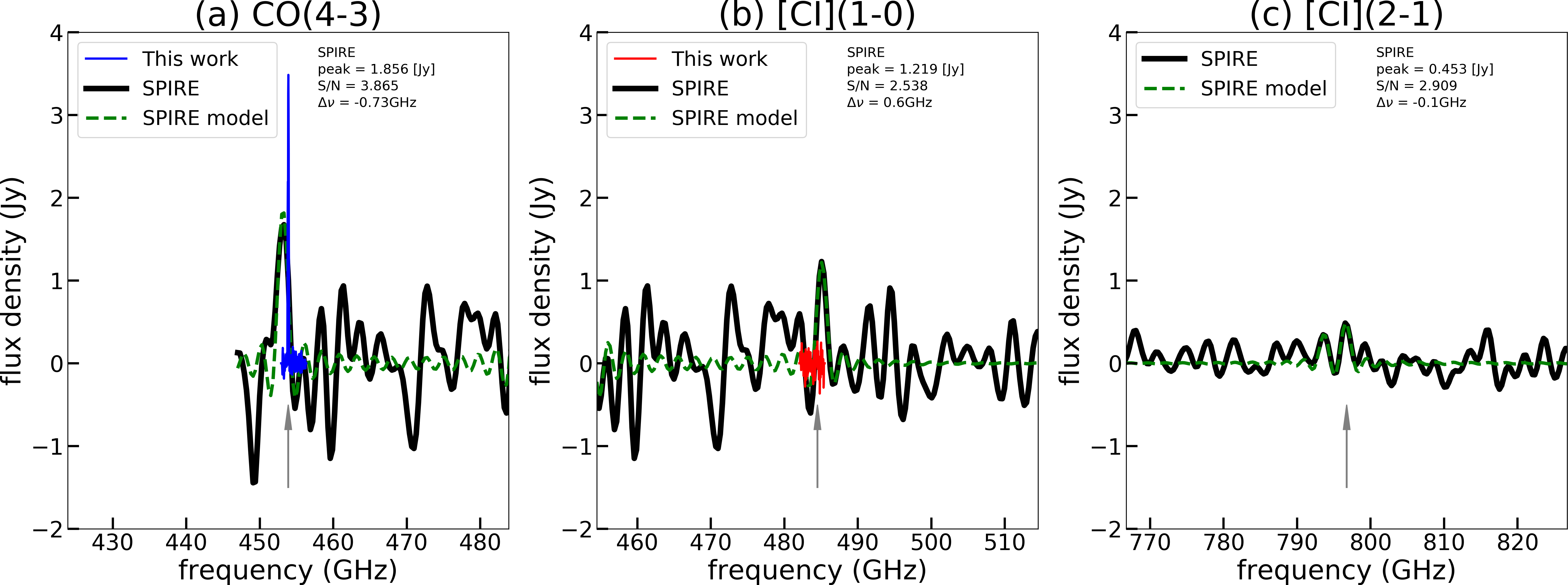}
\caption{(a)-(c) The black line indicates the continuum subtracted SPIRE/FTS spectrum around CO~(4--3), [\ion{C}{1}]~(1--0) and [\ion{C}{1}]~(2--1) emission. 
The best fit is shown as green dashed line.
The peak flux density and the frequency offsets from the systematic velocity measured by CO~(4--3) are shown at the right upper corners for each panel. The S/N is calculated by our own definition (see SECTION~\ref{sec:Appendix}).
Supplementary, the CO~(4--3) (blue) and [\ion{C}{1}]~(1--0) (red) spectrum obtained by this project are shown in (a) and (b), respectively.
The grey arrows indicate the frequencies of the CO~(4--3) and [\ion{C}{1}]~(1--0), and [\ion{C}{1}]~(2--1) corresponding to the systemic velocity. The second peak fitted in (c) is for CO~(7--6) emission line.
\label{fig:SPIRE}}
\end{center}
\end{figure}

NGC~6052 was observed by SPIRE/FTS \citep{Kame_2016, Lu_2017}.
We downloaded the science data products automatically generated by the data processing pipelines from ``herschel science archive \footnote{\url{http://archives.esac.esa.int/hsa/whsa/}}".
Same as \citet{Lu_2017}, we adopt point-source calibration and fit simultaneously the continuum and emission lines using a 5th order polynomial and sinc profiles following the ``Spectrometer Line Fitting" script in Herschel Interactive Data Processing Environment ({\tt HIPE}).
Figure~\ref{fig:SPIRE} shows the continuum subtracted SPIRE/FTS spectrum (black solid lines) and the fitting results (green dashed lines).
Only CO~(4-3) is marginally detected with the S/N of 3.9 and both [\ion{C}{1}]~(1--0) and [\ion{C}{1}]~(2--1) are not detected (S/N$\lesssim$3). In our analysis, S/N is calculated using the ratio between peak flux density obtained by spectrum fitting and the rms value of the flux density in the line free frequency range (i.e., 460-480~GHz for CO(4--3) and [\ion{C}{1}]~(1--0) and 800-820~GHz for [\ion{C}{1}]~(2--1) emissions), which is not the same method with \citet{Kame_2016} and \citet{Lu_2017}. 
Since the S/N is $\sim$3 for [\ion{C}{1}]~(1-0) and [\ion{C}{1}]~(2-1), these lines may be considered as ``detection" in different methods. For example, [\ion{C}{1}]~(1-0) emission line is considered as ``detection" in the analysis by \citet{Kame_2016}, but ``non-detection" by \citet{Lu_2017} and our analysis.
In Table~\ref{tab:SPIRE}, the velocity integrated flux (and $3\sigma$ upper limits) measured by our analysis, \citet{Kame_2016}, and \citet{Lu_2017} are shown.
Comparison of this SPIRE measurements with our Morita Array result implies that the CO~(4--3) flux recovered by the Morita Array may be $\sim30\%$ of the total.
But, given the fact that the {\em Herschel} beam is four times the solid angle of our FoV and the considerable uncertainties in the FTS measurement 
(i.e., The intensity is an analytical solution assuming sinc function for the spectroscopically unresolved line.\footnote{\url{http://herschel.esac.esa.int/hcss-doc-15.0/load/spire_drg/html/spire_spec_analysis.html}}), it is impossible to directly compare SPIRE flux and our Morita Array's results.
Further discussion of SPIRE/FTS measurements is beyond the aim of this letter because the line detection is marginal even for the CO~(4--3) emission.

\begin{deluxetable*}{cccc}
%\tablenum{1}
\tablecaption{SPIRE/FTS summary for NGC~6052 \label{tab:SPIRE}}
\tablewidth{0pt}
\tablehead{
\colhead{Line} & \colhead{$S_{\rm line}$d$v$} &  \colhead{$S_{\rm line}$d$v$} &  \colhead{$S_{\rm line}$d$v$} \\
\colhead{} & \colhead{Jy~km~s$^{-1}$} & \colhead{Jy~km~s$^{-1}$} & \colhead{Jy~km~s$^{-1}$} \\
\colhead{} & \colhead{our analysis} & \colhead{\citet{Kame_2016}} & \colhead{\citet{Lu_2017}
}
}
\decimalcolnumbers
\startdata
CO(4--3) & $1451\pm128$ & $1400^{+500}_{-400}$ & $1600\pm300$ \\
$[$\ion{C}{1}$]$~(1--0) & $<1055$ & $940^{+300}_{-310}$ & $<1058$ \\
$[$\ion{C}{1}$]$~(2--1) & $<208$ & $160^{+60}_{-72}$ & $210\pm30$
\enddata
\tablecomments{(1) Line name. (2)-(4) The velocity integrated line flux measured by this project, \citet{Kame_2016}, and \citet{Lu_2017}.}
\end{deluxetable*}

%\bibliography{sample63}{}

\begin{thebibliography}{}
%\bibitem[Alaghband-Zadeh et al.(2013)]{Alaghband-Zadeh_2013} Alaghband-Zadeh, S., Chapman, S.~C., Swinbank, A.~M., et al.\ 2013, \mnras, 435, 1493
\bibitem[Albrecht et al.(2007)]{Albrecht_2007} Albrecht, M., Kr{\"u}gel, E., \& Chini, R.\ 2007, \aap, 462, 575
\bibitem[Bolatto et al.(2013)]{Bolatto_2013} Bolatto, A.~D., Wolfire, M., \& Leroy, A.~K.\ 2013, \araa, 51, 207
\bibitem[Bothwell et al.(2017)]{Bothwell_2017} Bothwell, M.~S., Aguirre, J.~E., Aravena, M., et al.\ 2017, \mnras, 466, 2825
\bibitem[Burton et al.(2015)]{Burton_2015} Burton, M.~G., Ashley, M.~C.~B., Braiding, C., et al.\ 2015, \apj, 811, 13
\bibitem[Cicone et al.(2018)]{Cicone_2018} Cicone, C., Severgnini, P., Papadopoulos, P.~P., et al.\ 2018, \apj, 863, 143
\bibitem[Crocker et al.(2019)]{Crocker_2019} Crocker, A.~F., Pellegrini, E., Smith, J.-D.~T., et al.\ 2019, \apj, 887, 105
\bibitem[D{\'\i}az-Santos et al.(2017)]{Diaz-Santos_2017} D{\'\i}az-Santos, T., Armus, L., Charmandaris, V., et al.\ 2017, \apj, 846, 32
\bibitem[Glover \& Clark(2016)]{Glover_2016} Glover, S.~C.~O., \& Clark, P.~C.\ 2016, \mnras, 456, 3596
\bibitem[Habing(1968)]{Habing_1968} Habing, H.~J.\ 1968, \bain, 19, 421
\bibitem[Goldsmith et al.(2015)]{Goldsmith_2015} Goldsmith, P.~F., Y{\i}ld{\i}z, U.~A., Langer, W.~D., et al.\ 2015, \apj, 814, 133
\bibitem[Hainline et al.(2004)]{Hainline_2004} Hainline, L.~J., Scoville, N.~Z., Yun, M.~S., et al.\ 2004, \apj, 609, 61
\bibitem[Heintz \& Watson(2020)]{Heintz_2020} Heintz, K.~E., \& Watson, D.\ 2020, \apjl, 889, L7
\bibitem[Ikeda et al.(2002)]{Ikeda_2002} Ikeda, M., Oka, T., Tatematsu, K., et al.\ 2002, \apjs, 139, 467
\bibitem[Israel \& Baas(2002)]{Israel_2002} Israel, F.~P., \& Baas, F.\ 2002, \aap, 383, 82
\bibitem[Israel et al.(2015)]{Israel_2015} Israel, F.~P., Rosenberg, M.~J.~F., \& van der Werf, P.\ 2015, \aap, 578, A95
\bibitem[Izumi et al.(2018)]{Izumi_2018} Izumi, T., Wada, K., Fukushige, R., et al.\ 2018, \apj, 867, 48
\bibitem[James et al.(2002)]{James_2002} James, A., Dunne, L., Eales, S., et al.\ 2002, \mnras, 335, 753
\bibitem[Jiao et al.(2019)]{Jiao_2019} Jiao, Q., Zhao, Y., Lu, N., et al.\ 2019, \apj, 880, 133
\bibitem[Jiao et al.(2017)]{Jiao_2017} Jiao, Q., Zhao, Y., Zhu, M., et al.\ 2017, \apjl, 840, L18
\bibitem[Kamenetzky et al.(2016)]{Kame_2016} Kamenetzky, J., Rangwala, N., Glenn, J., et al.\ 2016, \apj, 829, 93
\bibitem[Kaufman et al.(2006)]{Kaufman_2006} Kaufman, M.~J., Wolfire, M.~G., \& Hollenbach, D.~J.\ 2006, \apj, 644, 283
\bibitem[Kaufman et al.(1999)]{Kaufman_1999} Kaufman, M.~J., Wolfire, M.~G., Hollenbach, D.~J., et al.\ 1999, \apj, 527, 795
\bibitem[Kramer et al.(2008)]{Kramer_2008} Kramer, C., Cubick, M., R{\"o}llig, M., et al.\ 2008, \aap, 477, 547
\bibitem[Krips et al.(2016)]{Krips_2016} Krips, M., Mart{\'\i}n, S., Sakamoto, K., et al.\ 2016, \aap, 592, L3
\bibitem[Lu et al.(2017)]{Lu_2017} Lu, N., Zhao, Y., D{\'\i}az-Santos, T., et al.\ 2017, \apjs, 230, 1
\bibitem[McMullin et al.(2007)]{McMullin_2007} McMullin, J.~P., Waters, B., Schiebel, D., et al.\ 2007, Astronomical Data Analysis Software and Systems XVI, 127
\bibitem[Miyamoto et al.(2018)]{Miyamoto_2018} Miyamoto, Y., Seta, M., Nakai, N., et al.\ 2018, \pasj, 70, L1
\bibitem[Mo et al.(2010)]{Mo_2010} Mo, H., van den Bosch, F.~C., \& White, S.\ 2010, Galaxy Formation and Evolution
\bibitem[Ojha et al.(2001)]{Ojha_2001} Ojha, R., Stark, A.~A., Hsieh, H.~H., et al.\ 2001, \apj, 548, 253
\bibitem[Oka et al.(2001)]{Oka_2001} Oka, T., Yamamoto, S., Iwata, M., et al.\ 2001, \apj, 558, 176
\bibitem[Papadopoulos et al.(2018)]{Papa_2018} Papadopoulos, P.~P., Bisbas, T.~G., \& Zhang, Z.-Y.\ 2018, \mnras, 478, 1716
\bibitem[Papadopoulos \& Greve(2004)]{Papa_2004_observation} Papadopoulos, P.~P., \& Greve, T.~R.\ 2004, \apjl, 615, L29
\bibitem[Papadopoulos et al.(2004)]{Papa_2004_theory} Papadopoulos, P.~P., Thi, W.-F., \& Viti, S.\ 2004, \mnras, 351, 147
\bibitem[Planck Collaboration et al.(2016)]{Planck_2016} Planck Collaboration, Ade, P.~A.~R., Aghanim, N., et al.\ 2016, \aap, 594, A13
\bibitem[Pound \& Wolfire(2008)]{Pound_2008} Pound, M.~W., \& Wolfire, M.~G.\ 2008, Astronomical Data Analysis Software and Systems XVII, 654
\bibitem[Rupke et al.(2008)]{Rupke_2008} Rupke, D.~S.~N., Veilleux, S., \& Baker, A.~J.\ 2008, \apj, 674, 172
\bibitem[Sage et al.(1993)]{Sage_1993} Sage, L.-J., Loose, H.-H., \& Salzer, J.~J.\ 1993, \aap, 273, 6
\bibitem[Salak et al.(2019)]{Salak_2019} Salak, D., Nakai, N., Seta, M., et al.\ 2019, \apj, 887, 143
\bibitem[Sanders et al.(2003)]{Sanders_2003} Sanders, D.~B., Mazzarella, J.~M., Kim, D.-C., et al.\ 2003, \aj, 126, 1607
\bibitem[Shi et al.(2005)]{Shi_2005} Shi, F., Kong, X., Li, C., et al.\ 2005, \aap, 437, 849
\bibitem[Shimajiri et al.(2013)]{Shimajiri_2013} Shimajiri, Y., Sakai, T., Tsukagoshi, T., et al.\ 2013, \apjl, 774, L20
\bibitem[Solomon \& Vanden Bout(2005)]{Solomon_2005} Solomon, P.~M., \& Vanden Bout, P.~A.\ 2005, \araa, 43, 677
\bibitem[Tadaki et al.(2018)]{Tadaki_2018} Tadaki, K., Iono, D., Yun, M.~S., et al.\ 2018, \nat, 560, 613
\bibitem[Taniguchi \& Noguchi(1991)]{Taniguchi_1991} Taniguchi, Y., \& Noguchi, M.\ 1991, \aj, 101, 1601
\bibitem[Tielens \& Hollenbach(1985)]{Tielens_1985} Tielens, A.~G.~G.~M., \& Hollenbach, D.\ 1985, \apj, 291, 722
\bibitem[Ueda et al.(2014)]{Ueda_2014} Ueda, J., Iono, D., Yun, M.~S., et al.\ 2014, \apjs, 214, 1
\bibitem[Valentino et al.(2018)]{Valentino_2018} Valentino, F., Magdis, G.~E., Daddi, E., et al.\ 2018, \apj, 869, 27
\bibitem[Valentino et al.(2020)]{Valentino_2020} Valentino, F., Magdis, G.~E., Daddi, E., et al.\ 2020, \apj, 890, 24
\bibitem[Walter et al.(2011)]{Walter_2011} Walter, F., Wei{\ss}, A., Downes, D., et al.\ 2011, \apj, 730, 18
\bibitem[Whelan et al.(2007)]{Whelan_2007} Whelan, D.~G., Devost, D., Charmandaris, V., et al.\ 2007, \apj, 666, 896
\end{thebibliography}
\bibliographystyle{aasjournal}



\end{document}